\documentclass[conference,american]{IEEEtran}

\pdfoutput=1

\usepackage{glossaries}

\usepackage{algorithm}
\usepackage{algpseudocode}
\usepackage{amsfonts}
\usepackage{tabularx}
\usepackage{tkz-euclide}
\usepackage{amsthm}
\usepackage{amssymb}
\usepackage{graphicx}

\newacronym{ecdf}{ECDF}{empirical cumulative distribution function}
\newacronym{tsnr}{TSNR}{transmit signal-to-noise ratio}
\newacronym{bcd}{BCD}{block coordinate descent}
\newacronym{bs}{BS}{base station}
\newacronym{ris}{RIS}{reconfigurable intelligent surface}
\newacronym{los}{LoS}{line-of-sight}
\newacronym{fcn}{FCN}{fully convolutional network}
\newacronym{snr}{SNR}{signal-to-noise ratio}
\newacronym{sinr}{SINR}{signal-to-interference-noise ratio}
\newacronym{ofdm}{OFDM}{orthogonal frequency-division multiplexing}
\newacronym{mmse}{MMSE}{minimum mean squared error}
\newacronym{mse}{MSE}{mean squared error}
\newacronym{mimo}{MIMO}{multiple-input-multiple-output}
\newacronym{wmmse}{WMMSE}{weighted minimum mean squared error}
\newacronym{noma}{NOMA}{nonorthogonal multiple access}
\newacronym{cnn}{CNN}{convolutional neural network}
\newacronym{drl}{DRL}{deep reinforcement learning}
\newacronym{imm}{IMM}{interacting multiple model}
\newacronym{ppo}{PPO}{proximal policy optimization}
\newacronym{sac}{SAC}{soft actor-critic}
\newacronym{ekf}{EKF}{extended Kalman filter}
\newacronym{gae}{GAE}{generalized advantage estimation}
\newacronym{rl}{RL}{reinforcement learning}
\newacronym{kl}{KL}{Kullback–Leibler}
\newacronym{rsu}{RSU}{road-side unit}
\newacronym{crlb}{CRLB}{Cram\'er-Rao lower bound}
\newacronym{aoi}{AoI}{age of information}
\newacronym{voi}{VoI}{value of information}
\newacronym{kf}{KF}{Kalman filter}
\newacronym{pdf}{PDF}{probability density function}
\newacronym{ao}{AO}{alternating optimization}
\newacronym{wsr}{WSR}{weighted sum-rate}
\newacronym{mrt}{MRT}{maximum ratio transmission}
\newacronym{zf}{ZF}{zero-forcing}
\newacronym{ml}{ML}{machine learning}
\newacronym{iid}{i.i.d.}{independent and identically distributed}
\newacronym{csi}{CSI}{channel state information}
\newacronym{sdma}{SDMA}{spatial division multiple access}
\newacronym{ue}{UE}{user equipment}
\newacronym{miso}{MISO}{multiple-input-single-output}
\newacronym{sc}{SC}{superposition coding}
\newacronym{sic}{SIC}{successive interference cancellation}
\newacronym{dpc}{DPC}{dirty paper coding}

\newcommand{\bs}[1]{\boldsymbol{#1}}

\newcommand{\bH}{\mathbf{H}}

\newcommand{\be}{\mathbf{e}}
\newcommand{\bh}{\mathbf{h}}

\newcommand{\bw}{\mathbf{w}}

\newcommand{\bj}{\mathbf{j}}

\newcommand{\bPhi}{\boldsymbol{\Phi}}

\newcommand{\E}{E}

\tikzstyle{feature} = [rectangle, minimum width=2cm, minimum height=.7cm, align=center, text centered, draw=black]
\tikzstyle{layer_short} = [rectangle, rounded corners, minimum width=3cm, minimum height=.7cm, align=center, text centered, draw=black]
\tikzstyle{layer} = [rectangle, rounded corners, minimum width=4cm, minimum height=.7cm, align=center, text centered, draw=black]

\newcommand{\norm}[1]{\lVert#1\rVert}

\usepackage{tikz}
\usetikzlibrary{spy}
\usetikzlibrary{shapes.geometric, arrows}
\tikzstyle{box} = [rectangle, rounded corners, minimum width=2cm, minimum height=0.8cm,text centered, draw=black,]
\usepackage{subfigure}
\usepackage{hhline}
\usepackage{pgfplots}
\usepackage{algpseudocode}
\usepackage{wasysym}
\usepackage{csquotes}
\pgfplotsset{compat=1.17}
\tikzstyle{box} = [rectangle, rounded corners, minimum width=2cm, minimum height=0.8cm,text centered, draw=black,]
\tikzstyle{label} = [minimum height=0.8cm,text centered, draw=none]

\usetikzlibrary{arrows,calc}
\tikzset{
>=stealth',
help lines/.style={dashed, thick},
axis/.style={<->},
important line/.style={ultra thick, smooth},
connection/.style={thick, dotted},
}

\tikzstyle{feature} = [rectangle, minimum width=2cm, minimum height=.6cm, align=center, text centered, draw=black, ]
\tikzstyle{layer} = [rectangle, rounded corners, minimum width=2.5cm, minimum height=.6cm, align=center, text centered, draw=black]
\usepackage{xparse}
\usetikzlibrary{3d}
\makeatletter % from https://tex.stackexchange.com/a/48776/121799

\usetikzlibrary{decorations.pathreplacing,matrix,shapes.arrows}

\newtheorem*{theorem*}{Theorem}

\usepackage[backend=biber, style=ieee, url=false, isbn=false]{biblatex}

\bibliography{ref.bib}

\pgfplotsset{compat=1.17}

\IEEEoverridecommandlockouts

\begin{document}

\title{Non-Orthogonal Multiple Access Assisted by Reconfigurable Intelligent Surface Using Unsupervised Machine Learning}
% Determin/Compute Phase Shifts of Reconfigurable Intelligent Surface for Non-Orthogonal Multiple Access with Unsupervised Machine Learning

\author{Finn Siegismund-Poschmann, Bile Peng and Eduard A. Jorswieck\\
\thanks{The work is supported by the Federal Ministry of Education and Research Germany (BMBF) as part of the 6G Research and Innovation Cluster 6G-RIC under Grant 16KISK031.}
\IEEEauthorblockA{Institute for Communications Technology, Technische Universit\"at Braunschweig, Germany\\Email: \{f.siegismund-poschmann, b.peng, e.jorswieck\}@tu-braunschweig.de}
}

\maketitle

\begin{abstract}
\Gls{noma} with multi-antenna \gls{bs} is a promising technology for next-generation wireless communication,
which has high potential in performance and user fairness.
Since the performance of \gls{noma} depends on the channel conditions,
we can combine \gls{noma} and \gls{ris},
which is a large and passive antenna array and can optimize the wireless channel.
However, the high dimensionality makes the \gls{ris} optimization a complicated problem.
In this work, we propose a machine learning approach to solve the problem of joint optimization of precoding and \gls{ris} configuration.
We apply the \gls{ris} to realize the quasi-degradation of the channel,
which allows for optimal precoding in closed form.
The neural network architecture RISnet is used,
which is designed dedicatedly for \gls{ris} optimization.
The proposed solution is superior to the works in the literature in terms of performance and computation time.
% It comprises many passive antennas,
% which reflect signals from the transmitter to the receiver with adjusted phases without changing the amplitude.
% The large number of the antennas enables a huge potential of signal processing despite the simple functionality of a single antenna.
% However, it also makes the \gls{ris} configuration a high dimensional problem,
% which might not have a closed-form solution and has a high complexity and, 
% as a result, 
% severe difficulty in online real-time application if we apply iterative numerical solutions.
% In this paper, we introduce a machine learning approach to maximize the \gls{wsr}.
% We propose a dedicated neural network architecture called \emph{RISnet}.
% The \gls{ris} optimization is designed according to the \gls{ris} property of product and direct channel and homogeneous \gls{ris} antennas.
% The architecture is scalable due to the fact that the number of trainable parameters is independent from the number of \gls{ris} antennas
% (because all antennas share the same parameters).
% The \gls{wmmse} precoding is applied and an \gls{ao} training procedure is designed.
% Testing results show that the proposed approach outperforms the state-of-the-art \gls{bcd} algorithm.
% Moreover, although the training takes several hours,
% online testing with trained model (application) is almost instant,
% which makes it feasible for real-time application.
% Compared to it, the \gls{bcd} algorithm requires much more convergence time.
% Therefore, the proposed method outperforms the state-of-the-art algorithm in both performance and complexity.
\end{abstract}

\begin{IEEEkeywords}
non-orthogonal multiple access,
reconfigurable intelligent surface, 
machine learning,
quasi-degradation.
\end{IEEEkeywords}

\glsresetall

% \fi
\IEEEpeerreviewmaketitle

\section{Introduction}
\label{sec:intro}

The \gls{noma} is a promising solution for future multiple access technique.
Unlike \gls{sdma},
which treats interference as noise,
\gls{noma} let users apply \gls{sic} to decode signals from the strongest one,
subtract it from the received signal,
until the desired signal of the user is decoded.
It has been shown that \gls{noma} has advantages in terms of spectral and energy efficiency as well as user fairness~\cite{maraqa2020survey}.

Compared to \gls{noma} with single-antenna \glspl{bs}~\cite{rezvani2021optimal},
precoding of \gls{noma} with multi-antenna \glspl{bs} has a higher potential of performance but also confronts new challenges.
It is proven that optimal precoding in a degraded multi-user \gls{miso} channel achieves the performance of the optimal \gls{sc} and \gls{sic}~\cite{Jorswieck_Rezvani_2021}.
The degraded channel, however, is rare in reality.
Therefore, the concept of \emph{quasi-degradation} is introduced.
A closed-form solution of optimal precoding is derived for quasi-degraded channels~\cite{Chen2016,Chen22016}.
Although the quasi-degradation is a relaxation compared to degradation,
this prerequisite is still a major challenge for the application.
As a solution, we propose to apply the \gls{ris} to optimize the channel.

\Gls{ris} is a large antenna array composed of many passive antennas.
It receives signals from the transmitter,
performs a simple signal processing without power amplification (e.g., phase shifting),
and transmits them to the receiver.
Due to the simple structure, low cost, and high integrability with other communication technologies,
the \gls{ris} is widely considered as a key enabling technology of the next-generation wireless communication systems~\cite{di2020smart,huang2020holographic}.
For decades, the channels were considered given and could not be modified.
The \gls{ris} enables a new opportunity to optimize the channels to become quasi-degraded~\cite{Zhu_Huang_Wang_Navaie_Ding_2021}.
Moreover, among the quasi-degraded channels,
more advantageous channels (e.g., with higher channel gains) are able to realize a higher performance.
These two considerations imply that the \gls{ris} can be a good company to \gls{noma}~\cite{Ding_Lv_Fang_Dobre_Karagiannidis_Al-Dhahir_Schober_Poor_2022}.
In the literature, optimization of multi-antenna \gls{noma} with an \gls{ris} has been performed with alternating difference-of-convex programming~\cite{Fu2019}, 
successive convex approximation~\cite{Zhu_Huang_Wang_Navaie_Ding_2021},
accurate and approximated closed-form solution~\cite{hou2020reconfigurable}
as well as machine learning~\cite{gao2021machine,liu2020ris,shehab2022deep,guo2022energy}. While the analytical methods \cite{Fu2019,Zhu_Huang_Wang_Navaie_Ding_2021,hou2020reconfigurable} are constrained by the suboptimality due to approximation and complexity,
the machine learning approaches~\cite{gao2021machine,liu2020ris,shehab2022deep,guo2022energy} have poor scalability (all the works listed here assume less than 100 \gls{ris} antennas,
which are far less than the vision of thousands of antennas~\cite{di2020smart}).

This work presents a joint optimization of precoding and \gls{ris} configuration with machine learning.
% We use the \gls{ris} to realize the quasi-degradation of channel and optimize channel further for better performance,
% while apply the optimal precoding for quasi-degraded channels.
The dedicated and scalable neural network architecture RISnet~\cite{peng2022risnet} is applied such that we can optimize a much larger \gls{ris} with up to 1024 antennas within a very short time because RISnet can parallelize the computation for each antenna.
We will show that the proposed solution is superior than the aforementioned works in terms of performance and computation time. % no performance yet -> better than random

\section{System Model and Problem Formulation}
\label{sec:problem}

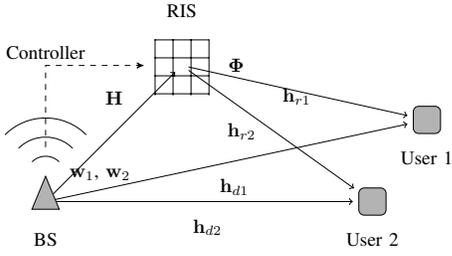
\begin{figure}[ht!]
    \centering
    \resizebox{.7\linewidth}{!}{    \begin{tikzpicture}
        \tikzstyle{base}=[isosceles triangle, draw, rotate=90, fill=gray!60, minimum size =.5cm]
        \tikzstyle{user}=[rectangle, draw, rotate=90, fill=gray!60, minimum size =.5cm, rounded corners=0.1cm]
        \tikzstyle{element}=[rectangle, fill=gray!30]
        
        \node[base] (BS) at (-3,0){};
        \draw[decoration=expanding waves,decorate] (BS) -- (-3,1.7);
        \node[user] (UE1) at (4,1.5){};
        \node[user] (UE2) at (3,0){};
        \draw[step=0.33cm] (-1,1.95) grid (0, 2.98);
        \node (RIS) at (-0.5, 2.5) {};
        
        \node[below of=BS,yshift=.3cm]{BS};
        \node[right of=BS,yshift=.5cm]{$\mathbf{w}_1$, $\mathbf{w}_2$};
        \node[below of=UE1,yshift=.3cm]{User 1};
        \node[below of=UE2,yshift=.3cm]{User 2};
        \node[above of=RIS]{RIS};
        \node[right of=RIS]{$\boldsymbol{\Phi}$};

        \draw[-to,shorten >=3pt] (BS) to node[below=2mm] {$\mathbf{h}_{d1}$} (UE1);
        \draw[-to,shorten >=3pt] (BS) to node[below=2mm] {$\mathbf{h}_{d2}$} (UE2);
        
        \draw[-to] (BS) to node[above=4mm] {$\mathbf{H}$} (RIS);
        
        \draw[-to,shorten >=3pt] (RIS) to node[left=1mm, pos=.6] {$\mathbf{h}_{r1}$} (UE1);
        \draw[-to,shorten >=3pt] (RIS) to node[below=-3mm, left=2mm] {$\mathbf{h}_{r2}$} (UE2);
        
        \draw[->, dashed] (-3, 1) |- node [above] {Controller} (-1.2, 2.5);
    \end{tikzpicture}}
    \caption{System model of RIS-assisted downlink broadcast channel.}
    \label{fig:system_model}
\end{figure}

The system model is an \gls{ris}-aided downlink scenario with multiple users.
In order to achieve a compromise between complexity and performance,
we assume two users in this work\footnote{An extension to more users is possible in two ways:
user clustering and higher order \gls{noma} processing.},
which is a widely applied assumption in the literature~\cite{Chen2016,Chen22016,Zhu_Huang_Wang_Navaie_Ding_2021}.  %todo: find more literature
The system model is shown in Fig.~\ref{fig:system_model}.

The \gls{bs} has $M$ antennas whereas the \gls{ris} is controlled by the \gls{bs} and has $N$ antennas (elements). 
The \glspl{ue} have one antenna each, which results in a multi-user \gls{miso} channel.

The received symbol $y_k$ of user $k$ is calculated as

\begin{equation}
    \label{eq:receivedsymb}
    y_k = \mathbf{h}_k^H \left( \mathbf{w}_1 x_1 + \mathbf{w}_2 x_2\right) + n_k, \quad k = 1,2,
\end{equation}
where $x_k \in \mathbb{C}$ is the transmitted symbol for user~$k$, 
$\E\left[\norm{x_k}^2 \right] = 1$, $\mathbf{w}_k \in \mathbb{C}^M$ is the precoding vector for user $k$ and $n_k \sim \mathcal{N}_{\mathbb{C}} (0, \sigma^2)$ is additive white Gaussian noise. 
The channel $\mathbf{h}_k \in \mathbb{C}^{M \times 1}$ is the sum of the channel via \gls{ris} and the direct channel, i.e.
\begin{equation}
    \mathbf{h}_k^H = \mathbf{h}_{rk}^H \bPhi \mathbf{H} + \mathbf{h}_{dk}^H, \quad k = 1,2,
    \label{eq:channel}
\end{equation}
where $\mathbf{h}_{dk}^H \in \mathbb{C}^{1 \times M}$ is the direct channel from \gls{bs} to user~$k$,
$\mathbf{h}_{rk}^H \in \mathbb{C}^{1 \times N}$ is the channel from \gls{ris} to user~$k$,
$\mathbf{H} \in \mathbb{C}^{N \times M}$ is the channel from \gls{bs} to \gls{ris},
$\boldsymbol{\Phi} \in \mathbb{C}^{N\times N}$ is the diagonal signal processing matrix of the \gls{ris}.
The diagonal elements $\phi_{nn}$ in row~$n$ and column~$n$ is $\phi_{nn} = e^{j\phi_n}$, which describes the phase shifts of the $n$th \gls{ris} antenna.

Without loss of generality, 
we assume that \gls{ue}~1 has the stronger channel gain and
\gls{ue}~2 has the weaker channel gain.
Following the \gls{sic} principle,
\gls{ue}~1 first decodes the stronger signal for \gls{ue}~2,
subtracts it from the received signal and decodes the signal for \gls{ue}~1 without interference,
whereas \gls{ue}~2 treats the signal for \gls{ue}~1 as interference and decodes the signal for \gls{ue}~2 directly.
\Gls{sinr} of signal for \gls{ue}~2 at \gls{ue}~1 $S_{21}$, 
\gls{snr} of signal for \gls{ue}~1 at \gls{ue}~1 $S_1$
and \gls{sinr} of signal for \gls{ue}~2 at \gls{ue}~2 $S_{22}$ are computed as
\begin{align}
    &S_{21} = \frac{\mathbf{h}_1^H \mathbf{w}_2 \mathbf{w}_2^H \mathbf{h}_1}{\mathbf{h}_1^H \mathbf{w}_1 \mathbf{w}_1^H \mathbf{h}_1 + \sigma^2}, \label{eq:SINR21}\\
    &S_{1} = \frac{\mathbf{h}_1^H \mathbf{w}_1 \mathbf{w}_1^H \mathbf{h}_1}{\sigma^2}, \label{eq:SNR1}\\
    &S_{22} = \frac{\mathbf{h}_2^H \mathbf{w}_2 \mathbf{w}_2^H \mathbf{h}_2}{\mathbf{h}_2^H \mathbf{w}_1 \mathbf{w}_1^H \mathbf{h}_2 + \sigma^2}, \label{eq:SINR22}
\end{align}
respectively.
% Since the signal for \gls{ue}~2 has to be decoded by both \gls{ue}~1 and \gls{ue}~2 and the signal for \gls{ue}~1 needs to be decoded only by \gls{ue}~1,
The achievable rates $R_1$ for \gls{ue}~1 and $R_2$ for \gls{ue}~2 are expressed as
\begin{align}
    &R_1 = \log \left(1 + S_1 \right), \label{eq:rate1}\\
    &R_2 = \min \left\{\log \left(1 + S_{21} \right), \log \left(1 + S_{22} \right)\right\}.\label{eq:rate2}
\end{align}

Our objective is to minimize the transmission power of the \gls{bs},
which is given by $||\mathbf{w}_1|| + ||\mathbf{w}_2||$,
by tuning the precoding vectors $\bw_1$, $\bw_2$ and the \gls{ris} configuration $\bPhi$,
subject to the required rates $r_1$ of \gls{ue}~1 and $r_2$ of \gls{ue}~2.
This problem can be formulated as\footnote{By assuming no maximum transmit power, the problem is always feasible.} \cite{Chen22016}
\begin{subequations}
\begin{alignat}{2}
&\!\min_{\mathbf{w}_1, \mathbf{w}_2, \bPhi} &\qquad& P=\norm{\mathbf{w}_1}^2 + \norm{\mathbf{w}_2}^2,   \label{eq:optProbMinPW}\\
&\text{subject to} &      & S_1 \geq 2^{r_1} - 1,  \label{eq:optProbMinPWconstraint1}\\
&                  &      & \min \left\{S_{21}, S_{22} \right\} \geq 2^{r_2} - 1,   \label{eq:optProbMinPWconstraint2}\\
&                  &      & |\phi_{nn}|=1,   \label{eq:diagonal_phi}\\
&                  &      & \phi_{nn'}=0 \text{ for } n \neq n'.   \label{eq:offdiagonal_phi}
% &                  &      & 0 \leq \psi_n \leq 2 \pi, \quad n=1,\dotsc,N.   \label{eq:optProbMinPWconstraint3}
\end{alignat}
\label{eq:problem}
\end{subequations}

\section{Optimal Precoding in Quasi-Degraded Channels}
\label{sec:quasi}

Given an \gls{ris} configuration $\bPhi$,
the optimization problem \eqref{eq:problem} with respect to $\bw_1$ and $\bw_2$ is not trivial to solve.
However, it is proved in \cite{Chen2016,Chen22016} that
there exists a closed-form optimal solution to $\bw_1$ and $\bw_2$
for quasi-degraded broadcast channels.
The broadcast channel is considered quasi-degraded if
\begin{equation}
    Q=\frac{1+r_1}{\cos^2 \psi} - \frac{r_1 \cos^2 \psi}{\left( 1 + r_2 \left(1 - \cos^2 \psi \right)\right)^2} \leq \frac{\norm{\bh_1}^2}{\norm{\bh_2}^2}
    \label{eq:quasi_degradation}
\end{equation}
where
\begin{equation}
    \cos \psi = \frac{\bh_1^H\bh_2\bh_2^H\bh_1}{\norm{\bh_1}\norm{\bh_2}}.
\end{equation}

In a quasi-degraded channel,
the optimal precoding vectors are obtained by~\cite{Chen22016}
\begin{align}
    \bw_1^* &= \alpha_1((1 + r_2)\be_1 - r_2\be_2^H\be_1\be_2)
    \label{eq:optimal_precoding1}\\
    \bw_2^* &= \alpha_2 \be_2
    \label{eq:optimal_precoding2}
\end{align}
where
\begin{align}
    \be_1 &= \frac{\bh_1}{\norm{\bh_1}},\\
    \be_2 &= \frac{\bh_2}{\norm{\bh_2}},\\
    \alpha_1^2 &= \frac{r_1}{\norm{\bh_1}^2}\frac{1}{(1 + r_2\sin^2\psi)^2},\\
    \alpha_2^2 &= \frac{r_2}{\norm{\bh_2}^2} + \frac{r_1}{\norm{\bh_1}^2}\frac{r_2\cos^2\psi}{(1 + r_2\sin^2\psi)^2}.
\end{align}

From the analysis above we can see that
the performance of \gls{noma} depends heavily on the channel for two reasons.
1) 
A prerequisite to apply the optimal precoding \eqref{eq:optimal_precoding1} and \eqref{eq:optimal_precoding2} is the quasi-degradation of the channel~\eqref{eq:quasi_degradation}.
2) Among the quasi-degraded channels,
more advantageous channels (e.g., with higher channel gains) are able to realize a lower transmission power subject to the rate requirements.
These two considerations imply that the \gls{ris} can be a good company to \gls{noma} because its ability to make a channel quasi-degraded and to optimize the quasi-degraded channel for a lower transmission power.
In Section~\ref{sec:ml},
we will introduce a machine learning approach that optimize the \gls{ris} configuration.

\section{Machine Learning Solution for Joint Precoding and RIS Configuration}
\label{sec:ml}

\subsection{Objective Function and Framework of Unsupervised Learning}
\label{sec:objective}

The objective function defines how the neural network is optimized.
It should 1) enforce the quasi-degradation~\eqref{eq:quasi_degradation} since it is the prerequisite of applying the optimal precoding~\eqref{eq:optimal_precoding1} and \eqref{eq:optimal_precoding2},
2) minimize the transmission power in the quasi-degraded channel.
The objective function is therefore formulated as
\begin{equation}
    L = \log \left(1 + \text{ReLU} \left(Q - \frac{\norm{\bh_1}^2}{\norm{\bh_2}^2} \right)\right) + \epsilon P,
    \label{eq:objective}
\end{equation}
where the first term is the penalty if the channel is not quasi-degraded
and the second term is the transmission power,
the constant factor $\epsilon > 0$ is chosen to balance the effort to make all channels quasi-degraded and to minimize the transmission power.

We define the neural network as $N_\theta$,
which is a function parameterized by $\theta$ and
maps from the channel feature $\boldsymbol{\Gamma}$,
which will be defined in Section~\ref{sec:feature_definition},
to the \gls{ris} phase shifts $\boldsymbol{\Phi}$,
i.e., $\boldsymbol{\Phi} = N_\theta(\boldsymbol{\Gamma}).$
With the optimal precoding for quasi-degraded channels presented in Section~\ref{sec:quasi},
our objective function $L$
is fully determined by the channel feature $\boldsymbol{\Gamma}$,
and the \gls{ris} configuration $\boldsymbol{\Phi}$.
We can write the objective as $ L(\boldsymbol{\Gamma}, \boldsymbol{\Phi})=L(\boldsymbol{\Gamma}, N_\theta(\boldsymbol{\Gamma}); \theta).$
Note that the right hand side of the equation emphasizes that $L$ depends on the parameter $\theta$ given $\boldsymbol{\Gamma}$.

We collect massive channel data in a training data set $\mathcal{D}$
and formulate the unsupervised machine learning problem as
\begin{equation}
    \min_\theta \sum_{\boldsymbol{\Gamma} \in \mathcal{D}} L(\boldsymbol{\Gamma}, N_\theta(\boldsymbol{\Gamma}); \theta).
    \label{eq:ml_objective}
\end{equation}
In this way, we optimize the function which maps from any $\boldsymbol{\Gamma} \in \mathcal{D}$ to $\boldsymbol{\Phi}$.
This optimization process is called \emph{training}.
If the data set is general enough,
we would expect that a channel feature $\boldsymbol{\Gamma}' \notin \mathcal{D}$,
which, however, is \gls{iid} as channel features in $\mathcal{D}$,
can also be mapped to a good \gls{ris} configuration.
The performance evaluation of $L(\boldsymbol{\Gamma}', N_\theta(\boldsymbol{\Gamma}'))$ for $\boldsymbol{\Gamma}' \notin \mathcal{D}$ and a trained and fixed $N_\theta$ is called \emph{testing}.

\subsection{Channel Features}
\label{sec:feature_definition}

We assume that $\mathbf{H}$ is constant because both \gls{bs} and \gls{ris} are fixed.
The neural network requires both $\bh_{dk}$ and $\bh_{rk}, k=1, 2$ to compute $\bPhi$.
The \gls{ris} optimization problem is complicated mainly because of the large number of \gls{ris} antennas.
However, the way that one single \gls{ris} antenna contributes to the overall channel is the same,
i.e., the \gls{ris} antennas are homogeneous.
Motivated by this fact,
we apply the same information processing for every \gls{ris} antenna in one layer of the multi-layer deep neural network.
This requires that the input of the deep neural network, i.e., the channel feature,
should be a stack of channel features per \gls{ris} antenna.
While this is straightforward for $\bh_{rk}^H$ 
because column~$n$ of $\bh_{rk}^H$ is the channel gain from \gls{ris} antenna~$n$ to user~$k$,
it is difficult for $\bh_{dk}^H$ since $\bh_{dk}^H$ is the direct channel from \gls{ris} to user~$k$.
Therefore, we apply the following trick:
\begin{equation}
    \bh_k^H=\bh_{rk}^H \bPhi \bH + \bh_{dk}^H = (\bh_{rk}^H \bPhi + \bj^H_k)\bH,
    \label{eq:}
\end{equation}
where we define $\bj^H_k = \bh_{dk}^H\bH^+$ with $\bH^+$ being the pseudo-inverse of $\bH$.
Column~$n$ of vector $\bj_k^H \in \mathbb{C}^{1\times N}$ can then be mapped to \gls{ris} antenna~$n$ unambiguously.
The channel feature $\boldsymbol{\Gamma}$ is defined as
\begin{equation}
\begin{aligned}
    \bs{\Gamma} = ( & |\bh^H_{r1}|; \arg(\bh^H_{r1}); |\bj^H_{1}|; \arg(\bj^H_{1})\\
    & |\bh^H_{r2}|; \arg(\bh^H_{r2}); |\bj^H_{2}|; \arg(\bj^H_{2})),
\end{aligned}
\label{eq:feature_definition}
\end{equation}
where the semicolon indicates a new row of the matrix.
$\bs{\Gamma}$ has a shape of $8\times N$.
Column~$n$ of $\bs{\Gamma}$ is the channel feature of \gls{ris} antenna~$n$.

\subsection{The RISnet Architecture}
\label{sec:architecture}

In this section, we present the RISnet architecture,
which is first introduced in~\cite{peng2022risnet}.
The basic idea of the RISnet is that an \gls{ris} antenna needs its local information as well as the information of the whole \gls{ris} to make a good decision on its configuration.
The local information of an antenna is obtained based on the information of the considered antenna only (therefore it is called local information).
The global information is the mean of the information of all \gls{ris} antennas,
which is the same to all \gls{ris} antennas and represents the information of the whole antenna array (therefore it is called global information).

Denote the input of layer~$i$ as $\mathbf{F}_i$ of shape $B_i \times N$, 
where $B_i$ is the feature dimension of layer~$i$ and the $n$th column of $\mathbf{F}_i$ is the feature vector of \gls{ris} antenna~$n$.
For $i = 1$, $\mathbf{F}_1^u$ (i.e., the input of the RISnet) is defined as the channel feature
\begin{equation}
    \mathbf{F}_1 = \boldsymbol{\Gamma}.
\end{equation}
% where the $n$th column of $\boldsymbol{\Gamma}$ is $\boldsymbol{\gamma}_n$ defined in \eqref{eq:feature_definition}.

For layer~$i < L$, we compute the local feature as
\begin{equation}
    \mathbf{F}^{l}_{i + 1} = \text{ReLU}(\mathbf{M}^l_i \mathbf{F}_i + \mathbf{b}_i^l)
    \label{eq:feature_local}
\end{equation}
where $\mathbf{M}^{l}_i$ of shape $B_{i+1}^l \times B_i$ and $\mathbf{b}_i^{l}$ of shape $B_{i+1}^l \times 1$ are trainable weight and bias for local feature in layer~$i$,
where $B_{i+1}^l$ is the local feature dimension for layer~$i+1$,
and ReLU is the rectified linear unit function.
Note that $\mathbf{b}_i^l$ is added to every column to $\mathbf{M}^l_i \mathbf{F}_i$.
The global feature is computed as
\begin{equation}
    \mathbf{F}^{g}_{i + 1} = \text{ReLU}(\mathbf{M}^{g}_i \mathbf{F}_i + \mathbf{b}_i^{g}) \times \mathbf{1}_{N} / N
    \label{eq:feature_global}
\end{equation}
for all \gls{ris} antennas, where $\mathbf{M}^{g}_i$ of shape $B_{i+1}^g \times B_i$ and $b_i^{g}$ of shape $B_{i+1}^g \times 1$ are trainable weight and bias for global feature in layer~$i$, with $B_{i+1}^g$ being the global feature dimension of layer~$i+1$,
and $\mathbf{1}_N$ is a matrix of all ones of shape $N\times N$.
% Compared to \eqref{eq:feature_local}, the global information is the same for all \gls{ris} antennas, which is the mean of $\text{ReLU}(\mathbf{W}^{g}_i \mathbf{F}_i + \mathbf{b}_i^{g})$ along the columns.
% Note that $\mathbf{b}_i^g$ is added to every column of $\mathbf{W}^{g}_i \mathbf{F}_i$ as in \eqref{eq:feature_local}.

The output feature by layer~$i$ is the concatenation of the channel features and the two features defined above:
\begin{equation}
    \mathbf{F}_{i + 1}^u = \left( 
    \left(\boldsymbol{\Gamma}\right)^T,  
    \left(\mathbf{F}_{i + 1}^{l}\right)^T,  
    \left(\mathbf{F}_{i + 1}^{g}\right)^T
    \right)^T.
\end{equation}
Therefore, the feature dimension of the layer~$i+1$ is $B_{i+1} = 8 + B_{i+1}^l + B_{i+1}^g$ since the channel feature dimension is $8$.

In the final layer ($i = L$), the output of the RISnet is computed as
\begin{equation}
    \mathbf{f}_{L + 1} = \text{ReLU}\left(\mathbf{m}_L \sum_u \mathbf{F}_{L}^u + b_L\right)
\end{equation}
where $\mathbf{m}_L$ and $b_L$ are trainable weights and bias, respectively. 
The \gls{ris} signal processing matrix $\boldsymbol{\Phi}$ is obtained by
\begin{equation}
    \boldsymbol{\Phi} = \text{diag}(e^{j\mathbf{f}_{L + 1}}).
    \label{eq:output2phi}
\end{equation}
Since elements in $\mathbf{f}_{L + 1}$ are always real,
we make sure the amplitudes of the diagonal elements in $\boldsymbol{\Phi}$ are 1 and the off-diagonal elements in $\boldsymbol{\Phi}$ are 0.

\begin{figure}
    \centering
    \resizebox{.95\linewidth}{!}{\begin{tikzpicture}
\node (channel) [feature] {$\boldsymbol{\Gamma}$};
\node (local) [feature, below of=channel, yshift=.3cm] {$\mathbf{F}_i^l$};
\node (global) [feature, below of=local, yshift=.3cm] {$\mathbf{F}_i^g$};

\draw [-stealth](-1.1,.5) -- node [anchor=south] {Antenna} (1,.5);
\draw [-stealth](-1.1,.5) -- node [anchor=east] {Feature} (-1.1,-1.8);

\draw [-](1.1,.35) -- (1.1,-1.75);

\node (local_layer) [layer_short, right of=local, xshift=2.5cm] {Local layer $i$};
\node (global_layer) [layer_short, right of=global, xshift=2.5cm] {Global layer $i$};

\draw [->] (1.1, -0.7) -- (local_layer.west);
\draw [->] (1.1, -0.7) -- (global_layer.west);

\node (channel2) [feature, right of=channel, xshift=6cm] {$\boldsymbol{\Gamma}$};
\node (local_output) [feature, below of=channel2, yshift=.3cm] {$\mathbf{F}_{i + 1}^l$};
\node (global_output) [feature, below of=local_output, yshift=.3cm] {$\mathbf{F}_{i + 1}^g$};

\draw [->] (local_layer.east) -- (local_output.west);
\draw [->] (global_layer.east) -- (global_output.west);
\end{tikzpicture}}
    \caption{Information processing of one layer in the RISnet.}
    \label{fig:info_proc}
\end{figure}
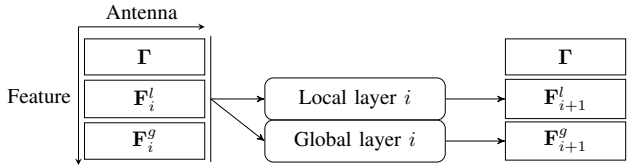

The information processing of one layer of the permutation-invariant RISnet is illustrated in Fig.~\ref{fig:info_proc}.
The RISnet has 8 layers and 8201 trainable parameters in total.
Compared to it,
a single layer of 8192 inputs and outputs (1024 antennas and 8 feature per antenna) has 67117056 parameters.
Training of the neural work is performed as Algorithm~\ref{alg:training} describes.

\begin{algorithm}
\caption{RISnet training}
\label{alg:training}
\begin{algorithmic}[1]
\State Randomly initialize RISnet.
\Repeat
\State Randomly select a batch of data samples.
\State Compute $\boldsymbol{\Phi}=N_\theta(\boldsymbol{\Gamma})$ for every data sample in the batch.
\State Compute the channel $\bh_k$ for $k=1, 2$ and every data sample in the batch.
\State Compute the objective function \eqref{eq:objective} for every data sample in the batch.
\State Compute the gradient of the objective w.r.t. the neural network parameters
\State Perform a stochastic gradient ascent step with the Adam optimizer
\Until{Predefined number of iterations achieved}
\end{algorithmic}
\end{algorithm}
\subsection{Complexity Analysis}
\label{sec:complexity}

% We need some statements about complexity/computation time

Due to the separation between offline training and online testing the online performance of the machine learning approach only depends on the size of the neural network. Obtaining the output of the trained RISnet requires only the computation of a forward propagation of the trained neural network~\cite{Matthiesen2020}. For this reason, the online complexity of the proposed  machine learning based method is much lower than the complexity of the SDR-based approaches from~\cite{Fu2019} which instead requires solving convex problems in each iteration.
\section{Training and Testing Results}
\label{sec:results}

\begin{figure}
    \centering
    \input{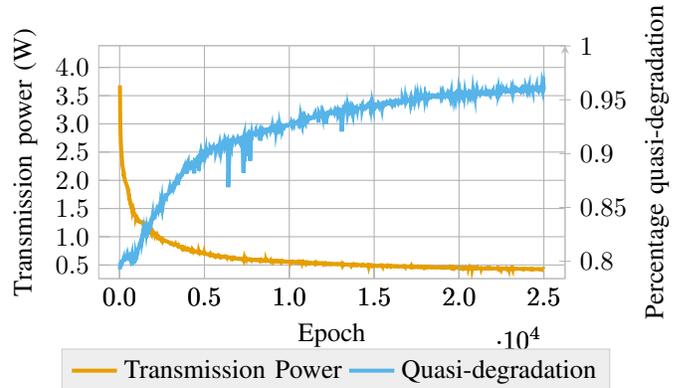}
    \caption{Training of the RISnet with $N = 1024$.}
    \label{fig:training}
\end{figure}

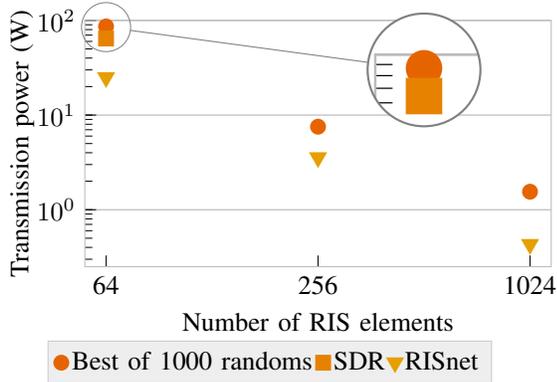
\begin{figure}
    \centering
    % This file was created with tikzplotlib v0.10.1.
\begin{tikzpicture}[spy using outlines={circle, magnification=2.3, connect spies}]

\definecolor{darkgray178}{RGB}{178,178,178}
\definecolor{lightgray204}{RGB}{204,204,204}
\definecolor{orange2301590}{RGB}{230,159,0}
\definecolor{orange2301300}{RGB}{230,130,0}
\definecolor{orange2301000}{RGB}{230,100,0}
\definecolor{silver188}{RGB}{188,188,188}
\definecolor{whitesmoke238}{RGB}{238,238,238}

\begin{axis}[
scale only axis, % The height and width argument only apply to the actual axis
height=\linewidth * 0.37,
width=\linewidth * 0.7,
axis line style={silver188},
legend cell align={left},
legend columns=3,
legend style={
  fill opacity=1,
  draw opacity=1,
  text opacity=1,
  at={(-0.08,-0.3)},
  anchor=north west,
  draw=lightgray204,
  fill=whitesmoke238
},
tick pos=left,
x grid style={white},
xlabel={Number of \gls{ris} elements},
xmajorgrids,
xmin=0.9, xmax=3.1,
xtick style={color=black},
xtick={0,1,2,3,4},
xticklabels={0,64,256,1024,4096},
y grid style={darkgray178},
ylabel={Transmission power (W)},
ymajorgrids,
ymin=-0.02945, ymax=100.0045,
ytick style={color=black},
ymode=log,
%ytick={-1,0,1,2,3,4,5,6,7,8},
%yticklabels={
%  \(\displaystyle {−1}\),
%  \(\displaystyle {0}\),
%  \(\displaystyle {1}\),
%  \(\displaystyle {2}\),
%  \(\displaystyle {3}\),
%  \(\displaystyle {4}\),
%  \(\displaystyle {5}\),
%  \(\displaystyle {6}\),
%  \(\displaystyle {7}\),
%  \(\displaystyle {8}\)
%},
y label style={yshift=-.2cm}
]
\addplot [thick, orange2301000, dotted, mark=*, mark size=2, only marks, line width=2pt, mark options={solid}]
table {%
1 86.51
2 7.54
3 1.55
};
\addlegendentry{Best of 1000 randoms}
\addplot [thick, only marks, orange2301300, mark=square*, mark size=2, line width=2pt, mark options={solid,rotate=180}]
table {%
1 64.45
};
\addlegendentry{SDR}
\addplot [thick, orange2301590, mark=triangle*, mark size=2, only marks, line width=2pt, mark options={solid,rotate=180}]
table {%
1 24.97
2 3.560
3 0.431
};
\addlegendentry{RISnet}
\coordinate (spypoint) at (1,85);
\coordinate (magnifyglass) at (2.5,30);
\end{axis}
\spy [gray, size=1.5cm] on (spypoint) in node[fill=white] at (magnifyglass);
\end{tikzpicture}
    \caption{Testing results of different approaches (SDR only works with 64 RIS elements).}
    \label{fig:testing}
\end{figure}

We apply the DeepMIMO framework to generate channel data~\cite{alkhateeb2019deepmimo}. We used the Outdoor 1 scenario with an intersection and placed the \gls{bs}, the \gls{ris} and the users so that there is a line of sight between the \gls{bs} and the \gls{ris} as well as between the \gls{ris} and the users, but not between the \gls{bs} and the users.
% The training and testing results are presented in this section.
Important parameters of scenario and model are presented in Table~\ref{tab:params}.
% The RISnet is trained to minimize the transmission power and increase the percentage of quasi-degraded channel paires according to the objective function \eqref{eq:objective}. 
The learning curve is shown in Fig.~\ref{fig:training}. The effect of the parameter $\epsilon$ is presented in Table~\ref{table:different_epsilon}. To estimate the performance the transmission power is compared to the best result of 1000 randomly generated phase shifts also using the optimal precoding for quasi-degraded channels. This simple baseline was used owing to the scalability up to 1024 \gls{ris}-elements. For 64 \gls{ris}-elements we also used an alternating SDR-based algorithm as described in \cite{Fu2019}. For larger numbers of \gls{ris}-elements this approach was not usable due to high memory consumption. The results are presented in Fig.~\ref{fig:testing}.
We can observe that the proposed method outperforms the baseline for all numbers of \gls{ris} antennas.
% The application of the RISnet on one channel feature is also significantly faster than the baseline on a laptop with Apple M2 Pro processing unit and 16~MB RAM.

\begin{table}[t]
    \centering
    \caption{Setting and Parameter values}
    \label{tab:params}
    \begin{tabular}{p{0.7\linewidth}p{0.25\linewidth}}
        \hhline{==}
        Parameter & Value \\
        \hline
        Number of \gls{bs} antennas & 9 \\
        Number of \gls{ris} antennas & \{64, 256, 1024\}\\
        % Number of users & 2\\
        % Channel models & Rayleigh fading channels\\
        % Transmit \gls{snr} & $10^{11}$\\
        % Weights of users & (0.25, 0.25, 0.25, 0.25)\\
        Number of layers & 8\\
        Learning rate & $5\times10^{-6}$\\
        % Optimizer & ADAM\\
        %Feature dimension for permutation-variant RISnet & 16\\
        Feature dimension & 8\\
        Iterations & 25000\\
        % Iterations for permutation-invariant version & 1000\\
        Batch size & 512\\
        % Optimizer & ADAM\\
        Number of data samples in training set & 10240\\
        Number of data samples in testing set & 1024\\
        \hhline{==}
    \end{tabular}
\end{table}

% The improvement of \gls{wsr} during training is shown in Fig.~\ref{fig:training},
% where one iteration is a gradient ascent step with a batch of channel data.
% From the figure we can observe that the training has improved the \gls{wsr} considerably.
% Furthermore, the permutation-variant RISnet quickly converges while the permutation-invariant RISnet takes significantly longer time to train.
% However, the achieved \gls{wsr} is higher with the permutation-invariant RISnet.
% The difference is bigger with higher \gls{tsnr}.
% It is to note that the permutation-invariant RISnet is only applicable to permutation-invariant objective functions.
% If the users have different weights
% or if we apply \gls{noma} instead of \gls{sdma},
% we have to use the permutation-variant RISnet
% because a user permutation changes the optimal \gls{ris} configuration.

\begin{table*}[ht!]
\centering
\caption{Training and test results for different values of $\epsilon$ with $N = 1024$.}
\vspace*{3mm}
\label{table:different_epsilon}
\resizebox{0.7\textwidth}{!}{
\begin{tabular}{|c | c c | c c | c c|} 
  \cline{2-7}
  \multicolumn{1}{c|}{} & \multicolumn{2}{c|}{$\epsilon = 1$} & \multicolumn{2}{c|}{$\epsilon = 0.1$} & \multicolumn{2}{c|}{$\epsilon = 0.01$} \\
 \cline{2-7}
  \multicolumn{1}{c|}{}  & Power & QD percentage & Power & QD percentage & Power & QD percentage \\
 \hline\hline
 Training & 0.325 & 91.6 \% & 0.421 & 96.1 \% & 2.353 & 96.7 \%\\ 
 \hline
 Test & 0.505 & 92.0 \% & 0.431 & 93.4 \% & 2.402 & 94.7 \% \\
 \hline
\end{tabular}}
\end{table*}

% We choose random phase shifts of the \gls{ris} and the \gls{bcd} algorithm proposed in~\cite{guo2020weighted} as two baselines to compare the proposed approach.
% The testing data are different from the training data and are identical to all four approaches shown in the figure.
% We can observe that both versions of the RISnet outperform the two baselines significantly.
% In addition, the testing with 1024 data samples takes only one minute on a laptop with RISnet.
% On the other hand, the optimization with the \gls{bcd} algorithm takes more than 24 hours on the server with the same data samples.
% This indicates that the proposed method is more advantageous in both performance and complexity,
% therefore is closer to reality.

\section{Conclusion}
\label{sec:conclusion}

We consider the joint optimization of \gls{noma} precoding and \gls{ris} optimization.
The precoding guarantees the optimal performance under the quasi-degraded channel constraint
and the \gls{ris} optimizes the channel to be quasi-degraded and to minimize the transmission power subject to the rate constraints.
The neural network architecture RISnet is applied to configure the \gls{ris},
which is designed dedicatedly for \gls{ris} optimization and its number of parameters is independent from the number of \gls{ris}-elements,
which makes it scalable.
We assume up to 1024 \gls{ris}-elements,
which are far more than in the assumptions in most literatures.
Testing results show good performance compared to the baseline and instant computation time.
Source code and data set are available under \url{https://github.com/bilepeng/risnet_noma}.

% We consider the \gls{wsr} maximization problem in an \gls{ris}-aided wireless communication network.
% Due to the complexity of the objective function,
% this problem does not have a closed-form solution.
% The large number of \gls{ris} antennas makes the optimization problem very high dimensional,
% which makes the computational complexity of an iterative numerical solution very high.
% In this work, we propose an unsupervised machine learning to solve this problem.
% A dedicated and scalable neural network architecture \emph{RISnet} is introduced,
% which achieves a high performance while retaining a low complexity.
% A permutation-variant version of the RISnet is introduced for general use cases and
% a permutation-invariant version is proposed for symmetric objective function,
% where the permutation of users does not change the RISnet output.
% This property makes it more stable to unseen channel data but limits its application to symmetric objective functions.
% Training and testing results show that the proposed solution achieves a better performance than the state-of-the-art algorithm with a much lower complexity in testing (application),
% thus making it not only more advantageous, but also closer to reality.

% \appendices

% \section{}

\ifCLASSOPTIONcaptionsoff
  \newpage
\fi


\printbibliography

\end{document}